\documentclass[10pt,conference,final]{IEEEtran}

\ifCLASSOPTIONcompsoc
  \usepackage[nocompress]{cite}
\else
  \usepackage{cite}
\fi
\ifCLASSINFOpdf
\else
\fi
\usepackage{booktabs}
\usepackage{hyperref}
\usepackage{listing}
\usepackage{booktabs}
\usepackage{graphicx}
\usepackage{paralist}
\usepackage{amssymb}
\usepackage{amsmath}
\usepackage{wrapfig}
\usepackage{color}
\usepackage[textsize=tiny]{todonotes}
\usepackage{bbding}
\usepackage{xcolor}
\usepackage{listings}
\usepackage{caption}
\usepackage[framemethod=tikz]{mdframed}
\definecolor{mycolor}{rgb}{0.122, 0.435, 0.698}
\newmdenv[innerlinewidth=0.5pt, roundcorner=4pt,linecolor=mycolor,innerleftmargin=2pt,
innerrightmargin=2pt,innertopmargin=2pt,innerbottommargin=2pt]{mybox2}

\usepackage{pdfpages}
\usepackage{array}
\usepackage[T1]{fontenc}
\usepackage{enumitem}

\newenvironment{normalize}{\leftskip-\leftmargin}{\par}

\mdfdefinestyle{mpdframe}{
    backgroundcolor            =lightgray!20,
    innerleftmargin            =4pt,
    innerrightmargin           =4pt,
    innertopmargin             =4pt,
    innerbottommargin           =4pt,
    linecolor                   =gray,
    skipabove                   =4pt
}

\newbool{revisionMode}
\boolfalse{revisionMode}

\newcommand{\revision}[2]{%
    \ifbool{revisionMode}
       {\color{black}{#1}}
       {\color{black}{#2}}%
    \color{black}
}

\newcommand{\materialurl}{\url{https://git.io/JtnTa} or an archived version at  \url{https://archive.softwareheritage.org/browse/revision/2e61f8ce57498194c2af0cd76e87498a174f07fa/}}

\newcolumntype{R}{>{$}r<{$}}
\newcolumntype{L}{>{$}l<{$}}
\newcolumntype{M}{R@{${}\cdot{}$}L}

\usepackage{color}

\begin{document}
%
\title{White-Box Performance-Influence Models:\\A Profiling and Learning Approach}

\author{\IEEEauthorblockN{Max Weber}
\IEEEauthorblockA{Leipzig University\\
Germany}
\and
\IEEEauthorblockN{Sven Apel}
\IEEEauthorblockA{Saarland University\\
Saarland Informatics Campus\\
Germany}
\and
\IEEEauthorblockN{Norbert Siegmund}
\IEEEauthorblockA{Leipzig University\\
Germany}}


%


\maketitle

\begin{abstract}
Many modern software systems are highly configurable, allowing the user to tune them for performance and more. Current performance modeling approaches aim at finding performance-optimal configurations by building performance models in a black-box manner. While these models provide accurate estimates, they cannot pinpoint causes of observed performance behavior to specific code regions. This does not only hinder system understanding, but it also complicates tracing the influence of configuration options to individual methods.

We propose a \emph{white-box approach} that models \emph{configuration-dependent performance behavior} at the \emph{method level}.
This allows us to predict the influence of configuration decisions on individual methods, supporting system understanding and performance debugging.
The approach consists of two steps: First, we use a coarse-grained profiler and learn performance-influence models for all methods, potentially identifying some methods that are highly configuration- and performance-sensitive, causing inaccurate predictions.
Second, we re-measure these methods with a fine-grained profiler and learn more accurate models, at higher cost, though.
By means of 9 real-world \textsc{Java} software systems, we demonstrate that our approach can efficiently identify configuration-relevant methods and learn accurate performance-influence models.
\end{abstract}

%



\section{Introduction}
Many software systems today are configurable, supporting multiple application scenarios, hardware platforms, and software stacks.
Configuration options are used to tailor a system's behavior and its non-functional properties by (de-)activating or tuning corresponding code.
Performance measures, such as response time and throughput, are among the most important non-functional properties of software systems~\cite{smith1993software, woodside2007future}.
So, it is crucial to know how individual configuration decisions will influence a system's performance.
Several approaches of accurately modeling and learning the performance behavior of configurable systems have been proposed in the literature~\cite{Siegmund:2013:FPM:2517208.2517209, siegmund2015performance, guo2013variability,GYS+18,SRK+12}.
The underlying idea is to sample a set of configurations from the configuration space and measure their performance.
Machine-learning techniques, such as multi-variable regression~\cite{SRK+12, sarkar2015cost} or classification  and  regression  trees~\cite{guo2013variability, GYS+18, sarkar2015cost} can be used to learn a performance-influence model from these measurements, to accurately predict the performance of unseen configurations or to find performance-optimal configurations with search-based techniques~\cite{Henard:2015, sayyad2013scalable, nair2018faster, oh2017finding}.

\begin{figure}
    \centering
    \includegraphics[width=0.98\linewidth]{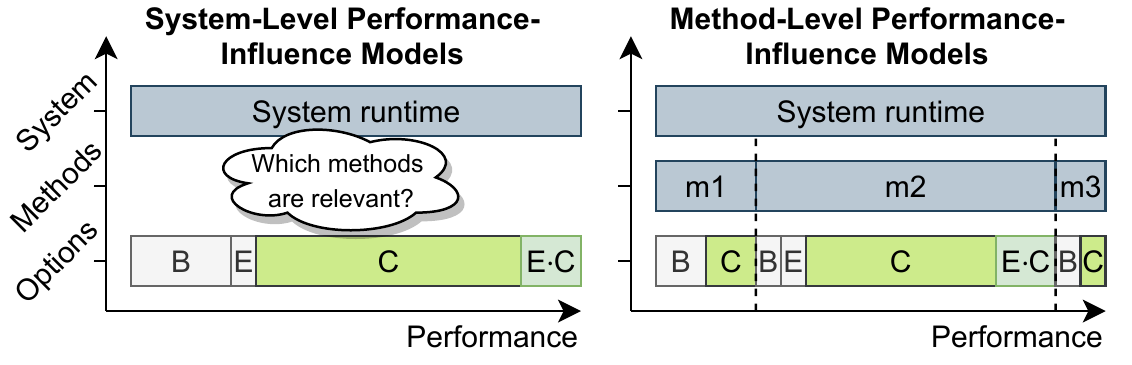}
    \caption{Comparison of the black-box performance-influence modeling at system level (left) and white-box performance-influence modeling at method level (right), explaining how feature values at system level are composed of feature values at method level.}
    \label{fig:wbpim}
\end{figure}

These and similar approaches based on parameter tuning~\cite{wang2018understanding} and algorithm selection~\cite{} have in common that they conceive the configurable software system as a \emph{black box}.
That is, they model the performance of a software system as a function of its configuration (i.e., a set of selected configuration options) without knowledge of the software's internals.
For illustration, let us consider a database system with three features \emph{Base} (\textsf{B}), \emph{Encryption} (\textsf{E}), \emph{Compression} (\textsf{C}), and \textsf{E$\cdot$C} (interaction between \textsf{E} and \textsf{C}), as illustrated in Figure~\ref{fig:wbpim} on the left side. From such a system-level model, we can infer the most influential options (e.g., \emph{Compression}) and possible interactions. However, we have no information about the root cause of interactions or influential options. We do not know where in the system's code base we spent execution time depending on the configuration. Essentially, developers want information at the level of individual methods, as illustrated in Figure~\ref{fig:wbpim} at right-hand side.

Knowing performance influences at the method level helps detecting performance bottlenecks~\cite{shen2015automating, ibidunmoye2015performance}, pinpointing performance bugs~\cite{jin2012understanding, han2016empirical}, or assigning performance tests in a CI pipeline to specific configurations.
This is a developer's perspective, which is not supported by current black-box approaches.

From a performance-analysis perspective, there are monitoring and profiling techniques whose goal is to identify performance hot spots~\cite{van2012kieker, selakovic2017actionable}.
Specialized performance engineers typically supervise performance of Web and cloud applications or identify bottlenecks using stress tests~\cite{snellman2011towards, chhetri2013smart}.
Unfortunately, state-of-the-art approaches in this area usually disregard the fact that today's software systems have huge configuration spaces.
Typically, only a single configuration of a system is considered, which is insufficient since performance bugs and related issues are often configuration-dependent~\cite{han2016empirical}.
Few approaches aim at creating white-box performance models of configurable systems~\cite{Siegmund:2013:FPM:2517208.2517209, velez2019configcrusher, velez2021WhiteBox}, but they either require explicit tracing information about which code regions are affected by which configuration options~\cite{Siegmund:2013:FPM:2517208.2517209} or they rely on expensive and potentially imprecise static and dynamic program analysis~\cite{velez2019configcrusher, velez2021WhiteBox}.

Our goal is to devise a white-box performance analysis technique for configurable software systems that, with high precision and at low cost, infers which methods are most affected by configuration decisions and which source-code regions exhibit the highest configuration-dependent performance variation.
Our approach substantially widens the application scenarios of former black-box approaches~\cite{Siegmund:2013:FPM:2517208.2517209, siegmund2015performance, guo2013variability} and introduces the concept of configurability and performance predictions to current white-box approaches~\cite{van2012kieker, selakovic2017actionable}.
The right part of Figure~\ref{fig:wbpim} shows the essence of the approach: it creates a performance-influence model for each method of the system.

To realize our goal, we combine approaches from two fields:
\begin{inparaenum}[(1)]
\item profiling program behavior in a white-box manner and
\item predictive modeling of performance of configurable software systems.
\end{inparaenum}
We use a two-step approach to direct the profiling activities to methods that are performance-relevant (i.e., contribute to a system's performance) and that are highly affected by configuration options.
This way, we substantially reduce the influence of measurement overhead, thereby increasing prediction accuracy.
In a nutshell, we draw samples from the configuration space based on well-established sampling strategies~\cite{medeiros2016comparison} and measure the selected system configuration with a low-overhead profiling tool.
We use classification and regression trees (CART) to learn a performance-influence model per method based on the measured performance data.
In this course, we identify methods that exhibit high performance variance, possibly caused by configuration options.
In a second step, we re-measure all methods with high variance to improve the accuracy of the final performance models, this way, focusing on the difficult-to-learn methods.
Here is the key: Methods are affected to different extents by configuration options, and only for a fraction of methods, a more fine-grained performance analysis is required. We use machine learning to find those methods.

Focusing on methods with high performance variation, we need to distinguish three causes: configuration variance, measurement variance, and a method's context variance.
\emph{Configuration variance} emerges from the (de-)selection of configuration options.
This is what we aim to learn as influences of options and interactions on individual methods.
\emph{Measurement variance} corresponds to the measurement setup's inherent systematic bias.
\emph{Context variance} represents the performance variation of a method execution due to varying input parameters or program states.
So, given these three types of variance, a method's performance can vary for each run of a software system, even with the same configuration.
To learn accurate performance models for configurable software systems, we need to measure, distinguish, and control sources for all three types of variance.

To provide a robust empirical foundation as a base line for evaluation, we measured the performance of 9 software systems from various application domains, resulting in 19 years of CPU time of continuous measurements. We demonstrate that our approach can efficiently identify configuration-relevant methods and learn accurate performance models at the method level.

We contribute and evaluate not only an approach for learning white-box performance models, but also important empirical findings about the distribution and nature of configuration-dependent performance-relevant methods.
We contribute an analysis that reveals the influence of the types of variance on the runtime of methods.
These findings shall inform further work on tailoring and guiding sampling techniques, static code analyses, and performance-anomaly detection.

To summarize, we make the following contributions:
\begin{compactitem}
\item An approach to learn white-box performance-influence models of configurable software systems at method level using profiling and prediction modeling;
\item A performance analysis providing insights into statistical performance properties of methods related to configuration decisions;
\item An evaluation of our approach with respect to prediction accuracy and scalability for 9 real-world software systems;
\item A replication package including our implementation and measurements\footnote{The supplementary material can be found at \materialurl}.
\end{compactitem}

\section{Preliminaries and Related Work}
\label{sec:related_work}

A black-box performance-influence model does not contain information to map performance to source code.
Yet, there exists substantial work conducting root-cause analysis, trying to locate, for example, performance bugs or memory leaks.
However, prior work rarely takes configurability into account.


\subsection{White-Box Performance Analysis}
\label{rel_work:wb_ana}
There are only few approaches that tackle performance analysis for configurable software systems.
Reisner et al.\ \cite{reisner2010using} and Meinicke et al.~\cite{meinicke2016essential} use symbolic and variational execution to analyze the behavior of interactions of configuration options at the level of control and data flow. They found that software systems in practice do have a much smaller relevant configuration space than theoretically possible, because only a few options interact at all. Many approaches in this area rely on this fact.

Hoffmann et al.\cite{Hoffmann:2011:DKR:1950365.1950390} use dynamic influence tracing to convert static parameters to dynamic control variables (global variables) with the goal of adapting properties of an application.
They do not consider interactions among parameters and cannot pinpoint code regions of interest.
Their approach works only when static parameters are convertible, whereas our approach is agnostic to the type of configuration parameters.

Family-based performance measurement~\cite{Siegmund:2013:FPM:2517208.2517209} aims at applying family-based analysis~\cite{thum2014classification} to performance analysis.
The idea is to create a variant simulator, which converts compile-time variability into run-time variability~\cite{von2016variability}. Then, the variant simulator is executed incorporating variability constraints. This way, multiple variants can be executed and measured in a single run by informing the analysis which method's performance is configuration-specific. On the downside, family-based performance measurement requires the construction of a variant simulator (or other variational representation), which is, in general, a non-trivial task~\cite{RLJ+18}.

Lillack et al.~\cite{lillack2018tracking} use taint analysis to identify which code fragments are executed depending on which configuration options. This static code analysis technique is used in Config-Crusher for deriving performance models~\cite{velez2019configcrusher}.
Specifically, ConfigCrusher employs static data-flow analysis to identify code regions whose performance is likely to be influenced by configuration decisions. It traces individual configuration options (represented by program variables)---following the call graph---and taints code regions influenced by configuration options or combinations thereof. Subsequently, ConfigCrusher merges regions to larger ones, such that it can efficiently measure performance of each individual region. Performance is measured by weaving instructions into the byte-code representation of the target program at the beginning and end of each region.

In contrast to our approach, ConfigCrusher requires modifying the source code of the target system to make the taint analysis run. Beside integrating an interface between analysis and target system, further substantial code refactorings are required to achieve scalability and precision~\cite{avdiienko2015mining}.
The background is that configuration options are often stored in complex data structures (lists, maps, structs, etc.), causing the taint analysis to no longer differentiate among the options stored in the data structures. The result is that either all accesses to the data structure are tainted with all configuration options or the analysis stops tainting at this point. The first variant leads to memory explosion and timeouts, the second results in incomplete and possibly very short taints, rendering the resulting performance models inaccurate and effectively useless.
To circumvent this problem, one can always refactor the entire system such that configuration options are stored in individual variables, which is usually infeasible in practice. Finally, in contrast to our approach, ConfigCrusher does not support numeric configuration options and is limited to single-threaded applications.

Velez et al.~\cite{velez2021WhiteBox} proposed \textsc{Comprex}, a tool that builds white-box performance models based on dynamic taint analysis and local performance measurement. \textsc{Comprex} requires expensive dynamic analysis and focuses only on configuration specific code, whereas our approach covers the whole code base by building a model for each method of the system.

\subsection{Black-Box Performance Analysis}
Obtaining accurate performance models requires a series of measurements. Conducting measurements for each possible configuration, however, is infeasible due to the combinatorial complexity of the problem. Instead, sampling a representative subset of configurations can achieve high accuracy. For configurable software systems, there are various sampling strategies that are suitable for learning performance models~\cite{medeiros2016comparison, kaltenecker2020interplay}: random sampling~\cite{guo2013variability, oh2017finding}, solver-based sampling~\cite{Henard:2015}, coverage-based sampling~\cite{johansen2012algorithm, Marijan:2013:PPT:2491627.2491646}, and  distance-based sampling~\cite{kaltenecker2019distance}.

Courtois and Woodside~\cite{courtois2000using} use regression splines to model the black-box performance behavior of a software system without taking configuration options into account. In the same vein, Israr et al.~\cite{israr2005automatic} and Mizan and Franks~\cite{mizan2011automatic} obtain performance models at a coarser granularity using Layered Queuing Networks without considering variability, though. Instead, they produce models that provide event sequences for distributed systems. Westermann et al.~\cite{westermann2012automated} aim at finding optimal software configurations with performance modeling at the black-box level. However, they consider only little variability. Also, Krogmann et al.~\cite{krogmann2010using} build parameterized performance models at the component level. That is, they build black-box models for components and do not address fine-grained, possibly cross-cutting configuration options among several components. Ackermann et al.~\cite{ackermann2018black} propose an approach to automatically find a suitable machine-learning technique to learn black-box performance models using monitoring data. Grohmann et al.~\cite{grohmann2019detecting} use feature selection in the context of machine learning to obtain black-box performance dependencies.
In contrast to these approaches, we use profiling information to build performance models at the method level, which is one of the main challenges and novelties of our approach.

Siegmund et al.~\cite{Siegmund2011scalable, siegmund2012predicting, siegmund2015performance} propose SPLConqueror, an approach to construct performance-influence models as linear functions over binary and numeric configuration options (or more complex combinations thereof). The key is to combine binary and numeric sampling and to symbolically learn the influence model in an iterative manner.
Other approaches propose learning techniques based on classification and regression trees~\cite{guo2013variability,GYS+18} and spectral learning~\cite{nair2018faster}, or even learn when to stop the learning procedure~\cite{sarkar2015cost}.
Our approach takes advantage of these modeling capabilities to pinpoint performance properties, but at the method level.

Nair et al.~\cite{nair2018finding} reduce the number of measurements by iteratively measuring and only adding configurations that improve accuracy the most.
Another approach by Nair et al.~\cite{nair2018faster} explores the configuration space by clustering, therefore, requiring measurements of only few representative configurations per cluster. This way, the sampling procedure can be directed to unveil performance influences and near-optimal configurations efficiently.
However, all these approaches consider a software system as a black box, not allowing for pinpointing performance behavior and root causes in code.

\subsection{Profiling}

Profiling refers to the white-box analysis of the run-time behavior of a program execution with respect to memory consumption or execution time~\cite{du2011performance}.
By contrast, measuring execution time of a system as a whole refers to black-box performance measurement.


Mytkowic et al.~\cite{Mytkowicz:2010:EAJ:1806596.1806618} analyzed the accuracy of Java profilers by comparing four commonly used profilers regarding their agreement on which methods are performance-critical.
The profilers reported different sets of methods as hot-spots.
Reasons for the disagreement are implementation details of the profilers (e.g., whether native methods are treated as part of the program) and the measurement overhead of the profiler (observer effect).

Some profiling approaches aim at automatically finding specific inefficient structures in the source code. Song and Lu~\cite{song2017performance} designed \textsc{LDoctor}, and Selakovic et al.~\cite{selakovic2017actionable} designed \textsc{DecisionProf}. Both tools search for redundant loops and optimization opportunities in the order of evaluating expressions. They focus only on specific code structures, but provide also suggestions for improvements.


There are many other approaches of how to profile different properties of software systems. However, these approaches do not consider configuration dependent performance variation. Still, profilers have been shown an important tool in industry to debug software systems with respect to resource usage. We do not want to replace but build on industry-strength profilers (i.e., \textsc{JProfiler}) to reach our goal.

\section{Untangling Performance Variance}
\label{sec:per_var}

At the method level, learning performance models is a task that is highly sensitive to measurement bias. As execution times can be short, influences of concurrent processes can easily distort measurements. In the context of configurability and method context, many sources contribute to the overall variation in performance. To use machine learning effectively on measurement data, we need to quantify possible sources of performance variance and adjust our model accordingly. To this end, we conducted an analysis of possible causes of performance variance: measurement variance, configuration variance, and context variance. To devise an approach for learning method-level performance models, it is necessary to know how the variance in the execution time is composed and how to control the three contributing factors to pin down the influence of configuration options on performance.

\begin{figure}
    \centering
    \includegraphics[width=\linewidth]{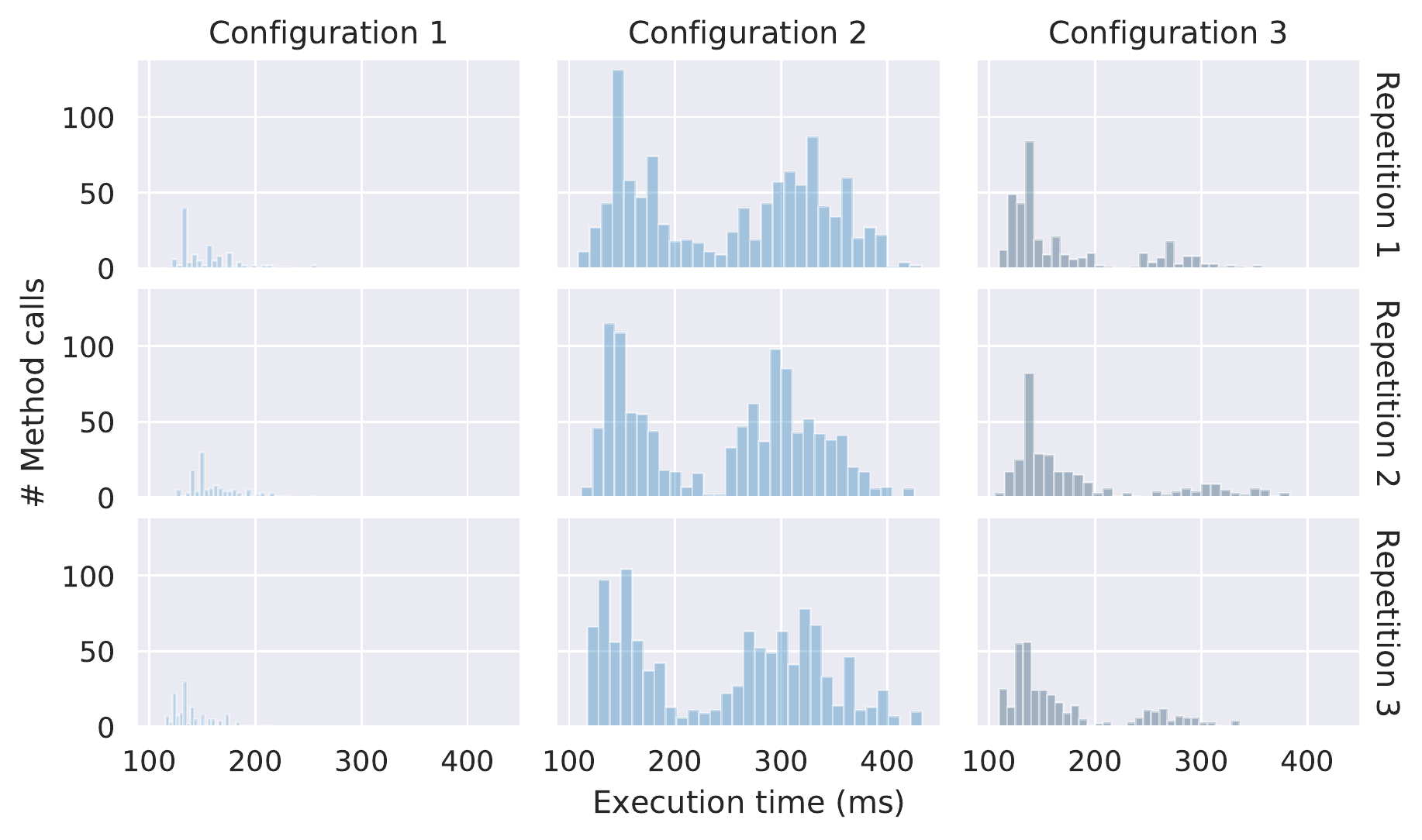}

    \caption{Performance variances for different executions of method \textsf{waitUntilSynced} of \textsc{Prevayler}. Columns represent different configurations (configuration variance), rows represent different repetitions of a program execution (measurement variance), and each cell depicts the performance distribution of multiple executions of the same method in a single program run (context variance).}
    \label{fig:var_overview}
\end{figure}

For illustration, Figure~\ref{fig:var_overview} shows the execution time of a single method executed in three different configurations, repeated three times each. Each histogram shows the distribution of the execution time of all calls to that method. {\em Measurement variance} becomes visible when comparing rows: The performance distribution of a row's plots should not change since the measurement setup is the same. That is, any change here can only be caused by the measurement process (e.g., overhead) or measurement environment (e.g., context switches). {\em Context variance} is represented by the shape of the histogram. That is, for different contexts during a single program run, we might call the method with different parameters, different cache states, etc., leading to different execution times. So, a histogram shows the distribution of execution times for a single method in a single program run. Finally, {\em configuration variance} manifests as differences among plots of different columns. If the plots differ across the columns, the root cause of the performance changes are due to changes in the system's configuration. In what follows, we provide an in-depth analysis of three real-world subject systems (\textsc{Catena}, \textsc{H2}, and \textsc{Prevayler}; see Section~\ref{sec:systems} for more details) regarding the three sources of variations. The goal is to obtain insights into which variance needs to be controlled when learning performance-influence models at method level. This helps us in devising sensible means for sampling, measuring, and learning instead of blindly applying an off-the-shelf machine-learning approach.


\subsection{Measurement Variance}
\label{sec:measurement_var}

Measurement variance affects the accuracy with which we get stable results while repeating experiments. High measurement variance adversely affects the accuracy of performance-influence models. Therefore, it is crucial to estimate measurement variance with a sufficient number of experiment repetitions.
The aim of our analysis is to determine the number of repetitions needed to trust the estimated measurement variance.

Our analysis setup is as follows: We profile a given configurable software system with 50 repetitions and report the coefficient of variation ($c_v = \frac{\sigma}{\mu}$; $\sigma$ represents the standard deviation of all method executions; $\mu$ the mean of all method executions in a single run) as a standardized measure of dispersion of a probability distribution to quantify the stability of measurement results~\cite{everitt1998cambridge}. To check whether measurement variance is independent of configuration variance, we randomly select 100 configurations for each of the three software systems. 

\begin{figure}
    \centering
    \includegraphics[width=0.98\linewidth]{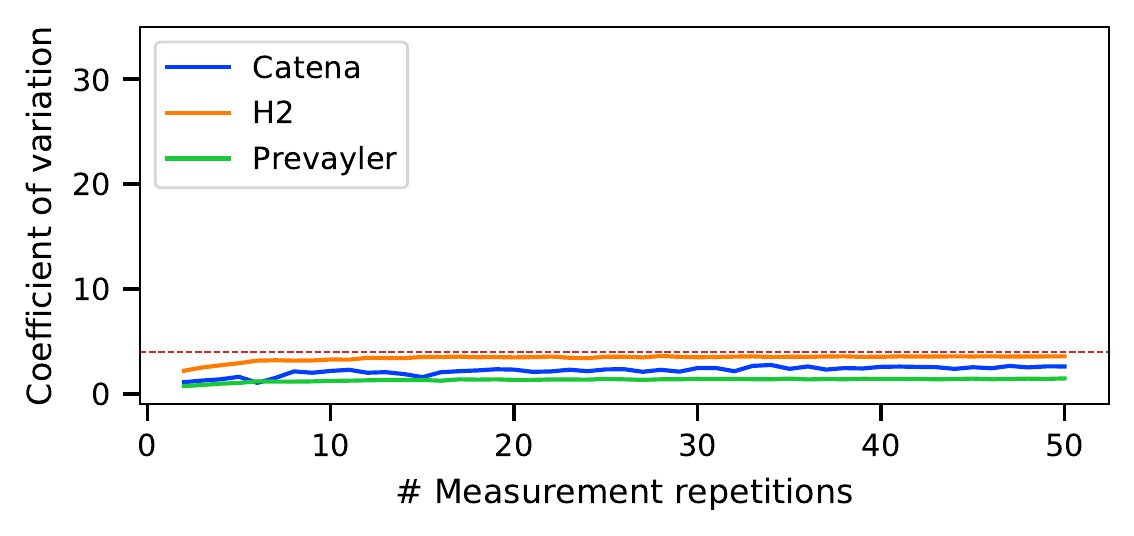}
    \caption{Measurement variance with increasing number of repetitions. The dashed red line denotes the maximal measurement variance (4\,\%).}
    \label{fig:var_rep}
\end{figure}

To better visualize the effect of repeating experiments, we compute the coefficient of variation after each repetition and show it for our three configurable software systems in Figure~\ref{fig:var_rep}.
For all three systems, the variance is below 5\,\%. This indicates a reliable measurement setup and \emph{the need for only a limited number of repetitions} (\textsc{Insight~1}) as the coefficient remains stable already around three repetitions. Moreover, the coefficient is equal for all configurations, so configurations have no effect on measurement bias. Hence, we can neglect possible hidden dependencies here.

\subsection{Configuration Variance}

Configuration variance captures the variation in a method's execution time due to selecting different configuration options.

Our analysis setup is as follows: We profile a given configurable software system by measuring the execution time of each method. We aggregate the execution time per method and repeat this process five times to account for measurement bias. We repeat this process for different configurations. Again, we use the coefficient of variation per method as a measure to determine whether methods have a constant average execution time across different configurations. If the coefficient of variation is higher than the measurement variance (4\,\% for all of our subject systems; see Section~\ref{sec:measurement_var}), the method's execution time is configuration-dependent.

\begin{figure}
    \centering
    \includegraphics[width=0.98\linewidth]{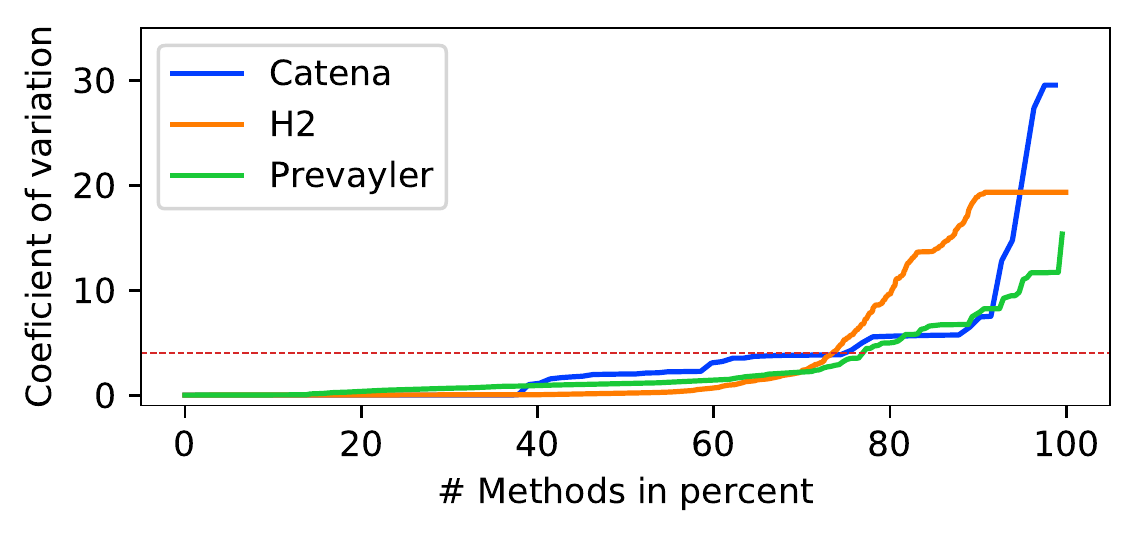}
    \caption{Configuration variance of the methods per subject system sorted by their coefficient of variation. The dashed red line denotes the maximal measurement variance (4\,\%) such that all method execution times above the line change due to configuration decisions.}
    \label{fig:var_conf}
\end{figure}

Figure~\ref{fig:var_conf} depicts the coefficient of variation (y-axis) for each method (x-axis) ordered from low variation to high variation. We observe that \emph{a large number of methods have only limited variance} (\textsc{Insight~2}) and \emph{only few methods exhibit high performance variance} (\textsc{Insight~3}). Interestingly, the variance for these few methods is huge, and the percentage of methods affected by configuration decisions is also sensitive to the software system. From this analysis, we infer that method-level performance-influence models should concentrate on these highly varying methods; profiling all methods of the program would be wasteful. This is good news as profiling is usually expensive and affects measurement results. Hence, we conclude that an efficient and accurate learning approach would first need to find the relevant methods and then concentrate learning these.

\subsection{Context Variance}

\label{sec:context_var}
Context variance of a method's execution time originates from changes in the method's calling context (e.g., method parameters and cache state). In Figure~\ref{fig:var_overview}, we visualize context variance as a histogram of performance values for each method execution in a single program run. We observe that the execution time of some methods remains constant during a program run, whereas other methods show high variance.

Overall, we observe \emph{highly skewed execution times that heavily affect a method's average execution time} (\textsc{Insight~4}). That is, we observe few but very large outliers, which are several orders of magnitude slower than about 99\,\% of the other method executions. This resembles a Cauchy distribution with no defined mean and standard deviation for these methods~\cite{pillai2016}. The problem for learning is that the Cauchy distribution is a well-known case where maximum-likelihood estimation fails and, subsequently, the likelihood principle in general~\cite{Ferguson78,reeds1985}. So, this can cause highly unreliable performance models.


Finally, we also see an interaction between context and configuration variance when analyzing Figure~\ref{fig:var_overview}: \emph{Not only the execution times vary, but also the number of method executions} (\textsc{Insight~5}), so an algorithm that takes only the average execution time of a method in to account for learning the influence of options is doomed to fail. Instead, an accurate approach needs to account for the \emph{number} of method calls in relation to their execution time.

\begin{figure}
    \centering
    \includegraphics[width=0.98\linewidth]{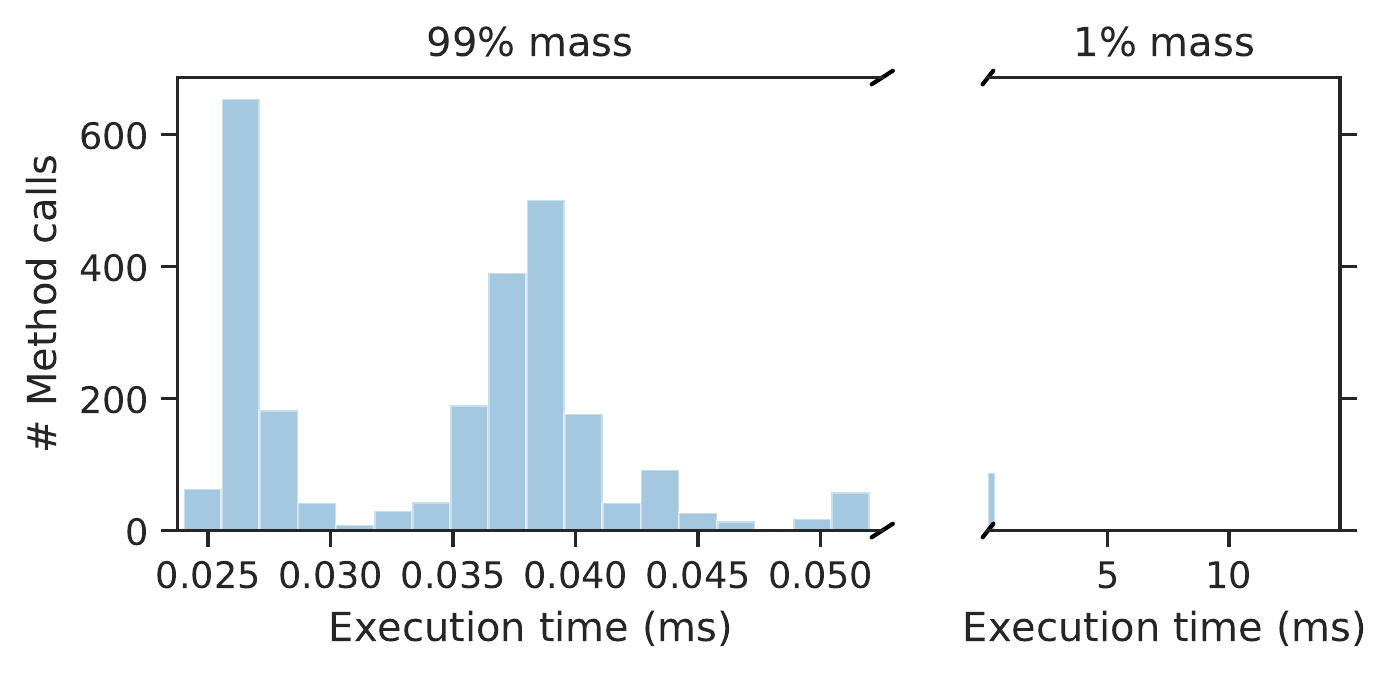}
    \caption{Context variance of method \textsf{rotr64}; right: the 1\,\% longest running method executions; left: the remaining 99\,\% method executions.}
    \label{fig:hist_filtered}
\end{figure}

\subsection{Summary}

From our analysis of variance, we can learn two important things: First, we see that the distribution of method executions changes for different configurations. That is, configuration influences a method's context causing a variation in the method's execution time. Second, based on our insight that some methods' performance values are Cauchy distributed, \emph{we cannot resort to a sample-based profiling technique, but rather need to tap the entire performance distribution using an instrumentation-based approach} (\textsc{Insight~6}). This is necessary as there might be large outliers that could skew the average method execution time substantially and would need to be filtered out. This is hardly possible with sample-based profiling.

\section{Performance-Influence Modeling at the Method Level}
\label{sec:white-box-monitoring}

\begin{figure*}
    \centering
    \includegraphics[width=0.85\textwidth]{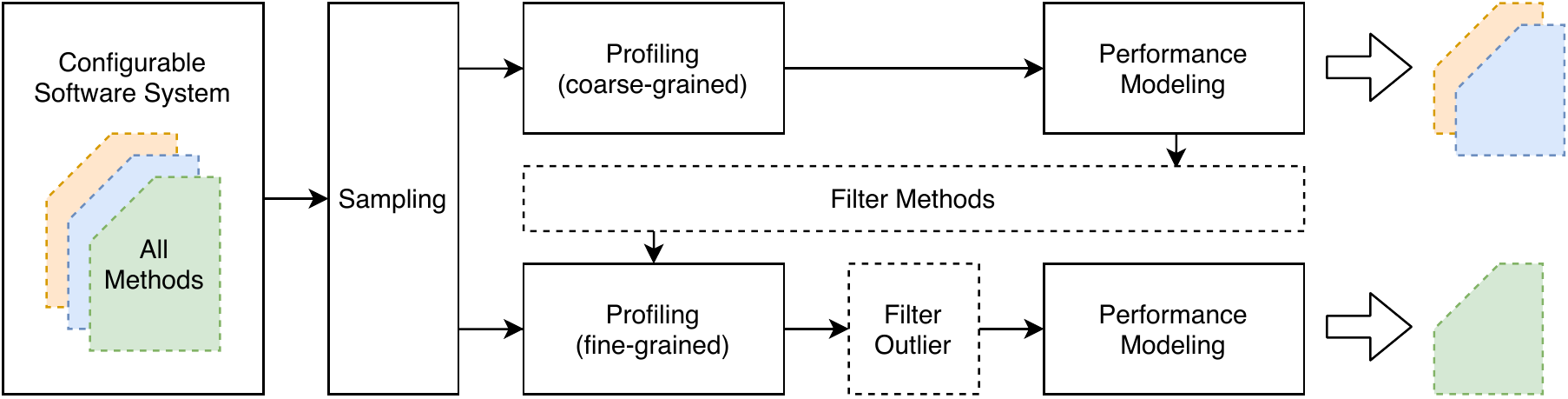}
    \caption{Method-level white-box-modeling pipeline for configurable software systems}
    \label{fig:workflowedge}
\end{figure*}

Our goal is to learn performance-influence models at the method level, so that we can pinpoint methods with high configuration-dependent performance variability and identify code regions that cause performance interactions.
Our method-level performance-influence modeling approach builds on the insights that we gained from our variance analysis and is separated into two steps, as illustrated in Figure~\ref{fig:workflowedge}: coarse-grained analysis and fine-grained analysis.

We know that the execution time of only few methods varies for different configurations (\textsc{Insight~3}), so we aim at identifying exactly these methods in the first step. For this purpose, we use a light-weight, coarse-grained profiler (\textsc{jProfiler}) to obtain performance measures for all methods under different configurations using an established sampling approach. Then, we extract those methods that exhibit (i) a performance-relevant execution time (e.g., we filter out getter/setter methods) and (ii) a performance variation across different configurations. In the second step, we instrument the source code (\textsc{Insight~6}) using the tool \textsc{Kieker} to obtain an execution time distribution (an execution time for each method call). We filter from this distribution long running outliers (\textsc{Insight~4}) and summarize the distribution as a histogram (\textsc{Insight~5}). Finally, we learn one performance-influence model per method based on these fine-grained values.


\subsection{Sampling Configurations}
\label{sec:sampling:cfgs}



As a prerequisite to step 1, we sample a set $\hat{C}$ from the set $C$ of all valid configurations. There is a substantial corpus of approaches that successfully applied different sampling strategies to obtain a representative set of configurations~\cite{oh2017finding, sarkar2015cost, kaltenecker2019distance, medeiros2016comparison}. Following previous work, we opt for feature-wise and pair-wise sampling for binary and Plackett-Burman sampling for numeric options~\cite{siegmund2015performance}, but other sampling strategies might be appropriate, as well.

With feature-wise sampling, we obtain a set of configurations in which each option is enabled once. With pair-wise sampling, also called t-wise sampling with t = 2, this set is enriched by all pair-wise combinations of configuration options. 
We use the extended Plackett-Burman design, proposed by Wang and Wu~\cite{wang1995hidden}, for sampling numeric configuration options. Compared to binary options, adding a numeric option with $n$ different values increases the configuration space by factor $n$ instead of factor 2. The Plackett-Burmann design selects a fixed set of configurations determined by a pre-chosen seed, which strongly reduces the effect of the combinatorial explosion. 

\subsection{Coarse-grained Profiling}
\label{sec:approach:cgp}
Our approach automatically runs a given software system for each configuration of the sample set (denoted as \emph{run}) with \textsc{jProfiler}~\footnote{\url{https://www.ej-technologies.com/products/jprofiler/overview.html}}, a coarse-grained profiler that uses the JVMTI interface of the JVM. For each run, we obtain the absolute execution time and the number of calls for each method. We repeat each run five times and report the mean time to account for measurement bias (\textsc{Insight~1}).

Next, we learn a performance-influence model per method $m \in M$ (where $M$ is the set of all methods of a software system) from these measurements using classification and regression trees (CART) as the learning method~\cite{guo2013variability}.
In the case that all methods have been learned accurately, there is no need to continue with the second step.
However, we have seen in our variance analysis that, typically, some methods are highly sensitive to context variance (\textsc{Insight~4 and~5}), which makes a second learning step necessary. 

\subsection{Filtering}
\label{sec:filtering}
To identify methods that are hard to learn and that contribute substantially to a system's performance (\textsc{Insight 5}), 
we apply a filter to all methods $M$ of a system obtaining a subset $M_{\mathit{hard}} \subseteq M$ for further measurement and learning (cf.~Eq.~\ref{eq:predicate}). 
The filter relies on a predicate $\phi(m, \alpha, \beta, \gamma)$, with $\alpha, \beta, \gamma \in \mathbb{R}$  (cf.~Eq.~\ref{eq:filter_final}), which states whether a given method $m \in M$ belongs to the set of performance-relevant methods.
\begin{equation}\label{eq:predicate}
    M_{hard} = \big\{\, m  \;|\;  m \in M \wedge \phi(m, \alpha, \beta, \gamma)\,  \big\}
\end{equation}
\begin{equation}\label{eq:filter_final}
    \dfrac{
        \Pi_{\mathit{err}}(m, \alpha) \qquad \Pi_{\mathit{abs}}(m, \beta) \lor \Pi_{\mathit{rel}}(m, \gamma)
    }{
        \phi(m, \alpha, \beta, \gamma)
    }
\end{equation}



$\Pi_{\mathit{err}}(m, \alpha)$ evaluates whether the error of the corresponding method's performance model exceeds the given threshold $\alpha$ (cf.~Eq.~\ref{eq:filter_a}). 
All methods that have been learned with a prediction error (mean absolute percentage error, MAPE, Eq.~\ref{eq:filter_a1}) of $\alpha$ or worse get selected. For this purpose, we compare the measured performance for method $m$ of configuration $c$ denoted with $\pi_c(m)$ with the performance $\pi'_c(m)$ predicted with the model of step 1. Following Siegmund et al.~\cite{siegmund2015performance}, we fix $\alpha$ to 5\,\% in our experiments, which represents an already strict filter criterion just above the measurement bias (cf. Section~\ref{sec:measurement_var}). Increasing $\alpha$ decreases the number of methods that have to be analyzed further, but this way more inaccurate performance models are accepted.
\begin{equation}\label{eq:filter_a}
\Pi_{\mathit{err}}(m, \alpha) = \text{MAPE}(m) \geq \alpha
\end{equation}
\begin{equation}\label{eq:filter_a1}
\text{MAPE}(m) = \frac{100}{|\hat{C}|} \cdot \sum_{c \in \hat{C}} \left\vert\frac{\pi_c(m) - \pi'_{c}(m)} {\pi_c(m)}\right\vert
\end{equation}

$\Pi_{\mathit{abs}}(m, \beta)$ (Eq.~\ref{eq:filter_b}) evaluates whether a method's execution time is longer than $\beta$. A method's execution time here is defined as the accumulated execution times over a run (Eq.~\ref{eq:filter_b1}). By setting $\beta$, we control to which extent we want to invest measurement effort for short-running methods.
\begin{equation}\label{eq:filter_b}
\Pi_{\mathit{abs}}(m, \beta) = \text{absPerf}(m) \geq \beta
\end{equation}
\begin{equation}\label{eq:filter_b1}
\text{absPerf}(m) = \frac{1}{|\hat{C}|} \cdot \sum_{c \in \hat{C}} \mathit{\pi}_{c}(m)
\end{equation}


$\Pi_{\mathit{rel}}(m, \gamma)$ (Eq.~\ref{eq:filter_c}) evaluates whether a method has a relative run-time (Eq.~\ref{eq:filter_c1}) of more than $\gamma$ in relation to the accumulated black-box time of the overall software system (sum of performance values of all methods, Eq.~\ref{eq:filter_c2}). Adjusting $\gamma$ enables us to focus on methods that contribute the most to the overall performance of the system.
\begin{equation}\label{eq:filter_c}
\Pi_{\mathit{rel}}(m, \gamma) = \text{relPerf}(m) \geq \gamma
\end{equation}
\begin{equation}\label{eq:filter_c1}
\text{relPerf}(m) = \frac{1}{|\hat{C}|} \cdot \sum_{c \in \hat{C}} \dfrac{\mathit{\pi}_{c}(m)}{\text{blackBoxPerf}(c)}
\end{equation}

\begin{equation}\label{eq:filter_c2}
\text{blackBoxPerf}(c) = \sum_{m \in M} \mathit{\pi}_{c}(m)
\end{equation}

To sum up, predicate $\phi$ selects methods that have been inaccurately learned ($\alpha$), have a total run time of, at least, $\beta$, and contribute to the overall software's performance by, at least, $\gamma$ percent.

\subsection{Fine-Grained Profiling and Learning}
We use \textsc{Kieker}~\cite{van2012kieker}, an aspect-oriented \textsc{Java} performance profiling tool, to measure the methods $M_{\mathit{hard}}$ obtained from the filtering step. To analyze the variation across all calls of a method, we extended \textsc{Kieker} by logging the assignment of values to method arguments of each method call together with the measured execution time.

Profiling with \textsc{Kieker} involves three steps:
First, we include an annotation (pointcut) into the source code of the subject system at the beginning of each selected method.
Second, we compile the software system into an executable.
Third, we execute the software with \textsc{Kieker} as JVM argument (\textsc{Java} agent) to weave the monitoring code (advice) around method executions.
We run our experiments with the same set of configurations and workloads as used in the coarse-grained profiling phase, obtaining performance data per method execution of the relevant methods. Based on our variance analysis, we filter outliers that make up 1\,\% of the longest execution times and we learn new models with the remaining aggregated data using CART.

\section{Experiment Setup}

In this section, we present the measurement setup that we use for profiling as well as the software systems that we selected for evaluation.

\subsection{Measurement Setup}

All measurements ran on a cluster of 27 computers, each of which has an Intel Quad-Core processor, an SSD running a headless operating system (Ubuntu 18.04.3 LTS), an HDD to store experiment data, and 8 or 16GB of RAM.\footnote{For a single software system, we conducted all measurements either on the systems with 8 or 16GB memory.}

\subsection{Measurement Procedure}
\label{sec:measurement:procedure}

For each subject system, we generate two sample sets for learning (according to the two sampling strategies of Section~\ref{sec:sampling:cfgs}), as shown in Table~\ref{tab:software_systems}. For each learning set, we measure the runtime with our two-step approach, learning a model per method. We sample an additional test set of 100 fresh configurations at random. We use the test set to evaluate the prediction error of the models learned based on the learning sets.
Furthermore, we repeat all measurements five times, which results in 94,000 measurement runs (RQ$_1$ and RQ$_2$). Additionally, we measure the black-box execution time of all configurations to determine the execution time without profiling (RQ$_3$).

\subsection{Subject Systems}\label{sec:systems}

We evaluate our approach with 9 real-world software systems. Our selection includes configurable \textsc{Java} applications, covering different domains, including databases, rendering engines, and static code analyzers. Our selection was driven by covering a diverse set of domains, having memory and CPU-intensive tasks, and providing configuration options that affect performance of the system. We provide an overview of the software systems in Table~\ref{tab:software_systems}. Next, we present the systems and benchmark workloads. When possible, we reused existing workloads provided by the respective software systems.~\footnote{We provide all benchmarks on the supplementary Website: \materialurl}

\begin{table}[t]

\caption{Overview of subject systems. $|F|$ denotes the number of configuration options (binary and numeric); $|C|$ denotes the number of the valid configurations $C$; $|\hat{C}_{\mathit{FW}}|$ denotes the number of configurations sampled feature-wise; $|\hat{C}_{\mathit{PW}}|$ denotes the number of configurations sampled in a pair-wise manner.}
\addtolength{\tabcolsep}{-3pt}
\begin{tabular}{llr r @{\,} c @{\,} r @{} l rr}
\toprule
    System & Domain & $|F|$ & \multicolumn{4}{c}{$|C|$}  & $|\hat{C}_{\mathit{FW}}|$ & $|\hat{C}_{\mathit{PW}}|$ \\
\midrule
    \textsc{Batik} & SVG rasterizer & 31 & $9.6$ & $\cdot$ & $10$  & $^{4}$ & 28 & 337 \\
    \textsc{Catena} &   Password hashing &   12 &  $1.0$ & $\cdot$ & $10$  & $^9$ & 875 & 2625 \\
    \textsc{cpd} &      Copy-paste detector &    7 &  $1.1$ & $\cdot$ & $10$  & $^4$ & 40 & 115 \\
    \textsc{dc} &  Image density converter &   24 &  $3.4$ & $\cdot$ & $10$  & $^6$ & 1600 & 9700 \\
    \textsc{H2} &                 Database &   16 & $6.5$ & $\cdot$ & $10$  & $^{11}$ & 375 & 2275 \\
    \textsc{Kanzi} & Data compression & 40 & $4.3$ & $\cdot$ & $10$  & $^{3}$ & 34 & 458 \\
    \textsc{pmd} &     Source-code analyzer &   11 &  $5.1$ & $\cdot$ & $10$  & $^2$ & 36 & 104 \\
    \textsc{Prevayler} &        Database &   12 &  $1.3$ & $\cdot$ & $10$  & $^5$ & 250 & 400 \\
    \textsc{Sunflow} &    Rendering engine &    6 &  $5.4$ & $\cdot$ & $10$  & $^6$ & 125 & n/a \\

\bottomrule
\end{tabular}

    \label{tab:software_systems}
\end{table}

The \textsc{Batik} rasterizer converts SVG files to a raster format. As workload, we used the \textsc{DaCapo} benchmark suite~\cite{blackburn2006dacapo}, which contains a set of SVG images of different sizes that can be used for performance tests.

\textsc{Catena} is a secure password scrambling framework that implements a corresponding hashing function. As workload, we used its password hashing benchmark. With the provided configuration options, it is possible to select one out of four graphs as well as different seeds and security values that influence how much main memory has to be used to encrypt or decrypt a password and how long this process takes.

\textsc{Cpd} is a code duplication detector. It detects duplicate source code sections to support developers with code refactoring. As a benchmark workload, we detect code duplicates in \textsc{Catena}'s code base.

\textsc{Density-converter} (DC) is an image density converter that, given an image or folder, converts these inputs into image formats with different resolutions. As workload, we used a set of high resolution images provided by the developers.

\textsc{H2} is an open-source relational database system that can operate both in an embedded and a client-server setting. As workload, we use the subset of tests of the \textsc{PolePosition} benchmark, with which developers compare \textsc{H2} to other database applications.

\textsc{Kanzi} is a lossless data compressor. It provides various configuration options for composing and tuning the compression process. As workload, we used the \textsc{Silesia} corpus benchmark~\footnote{Silesia Corpus: \url{http://sun.aei.polsl.pl/~sdeor/index.php?page=silesia}}.

\textsc{Pmd} is an extensible cross-language static code analyzer that checks source code against a set of rules. As a workload, we selected all rules that try to identify performance violations in the system to analyze. The system that we analyzed is \textsc{Prevayler}.

\textsc{Prevayler} is an open-source object persistence library for \textsc{Java} supporting in-memory storage. It provides a scalability and performance benchmark consisting of transaction-processing and query scalability tests that are applied to a JDBC-compatible database.

\textsc{Sunflow} is an open-source global illumination rendering system. It provides a selection of example scenes (objects to illuminate and render), of which we selected the \textit{golden} scene (a teapot in a colored room with one light source and 128\,$\times$\,128 pixels).

\section{Evaluation}

The goal of our approach is to pin down the influence of configuration options on individual methods. In our evaluation, we address three research questions:









\begin{itemize}[leftmargin=2.7em]

    \item[{\bfseries RQ$_1$:}] Can we learn accurate performance-influence models at the method level?

\begin{normalize}
Previous work has shown that performance can be accurately modeled for a system as a whole. However, it is unclear whether this level of accuracy can be achieved when modeling performance at the method level, due to measurement overhead and the various sources of variance that we have described in Section~\ref{sec:per_var}.
\end{normalize}

    \vskip 1pt
    \item[{\bfseries RQ$_2$}:] How do system-level and method-level models compare in terms of information they provide?

\begin{normalize}
Knowing the influence of a configuration option at the system level is helpful for tuning a system's performance. However, identifying the root cause of the influence of features helps developers, for example, to spot performance bugs and to focus on specific performance tests in a CI pipeline.
\end{normalize}

    \vskip 1pt
    \item[{\bfseries RQ$_3$}:] What is the relation of the runtimes of  profiled and unprofiled methods?

\begin{normalize}
An important measure for validity is whether the actual (unprofiled) method execution time relates to the measured (profiled) method execution time. Profiling introduces overhead, and therefore the models we learn may be biased. With this question, we aim at quantifying the extent of the profiling overhead.
\end{normalize}

\end{itemize}

As ConfigCrusher is closest to our approach (cf.~Section~\ref{sec:related_work}), it would be a natural candidate for comparison. In Section~\ref{sec:eval:cc}, we report on why a comparison is not feasible, though.

\subsection{Method-Level Performance Models ($RQ_1$)}

\paragraph{Operationalization}

To answer RQ$_1$, we calculate the MAPE of the performance-influence models of each method separately. We follow the measurement procedure described in Section~\ref{sec:measurement:procedure}. We assume an influence model to be sufficiently accurate if it predicts the execution time of the method in the test set with an average error below 5\,\%, which is stricter than 10\,\% as used in previous work~\cite{guo2013variability}.

\paragraph{Results}

\begin{figure}
    \centering
    \includegraphics[width=\linewidth]{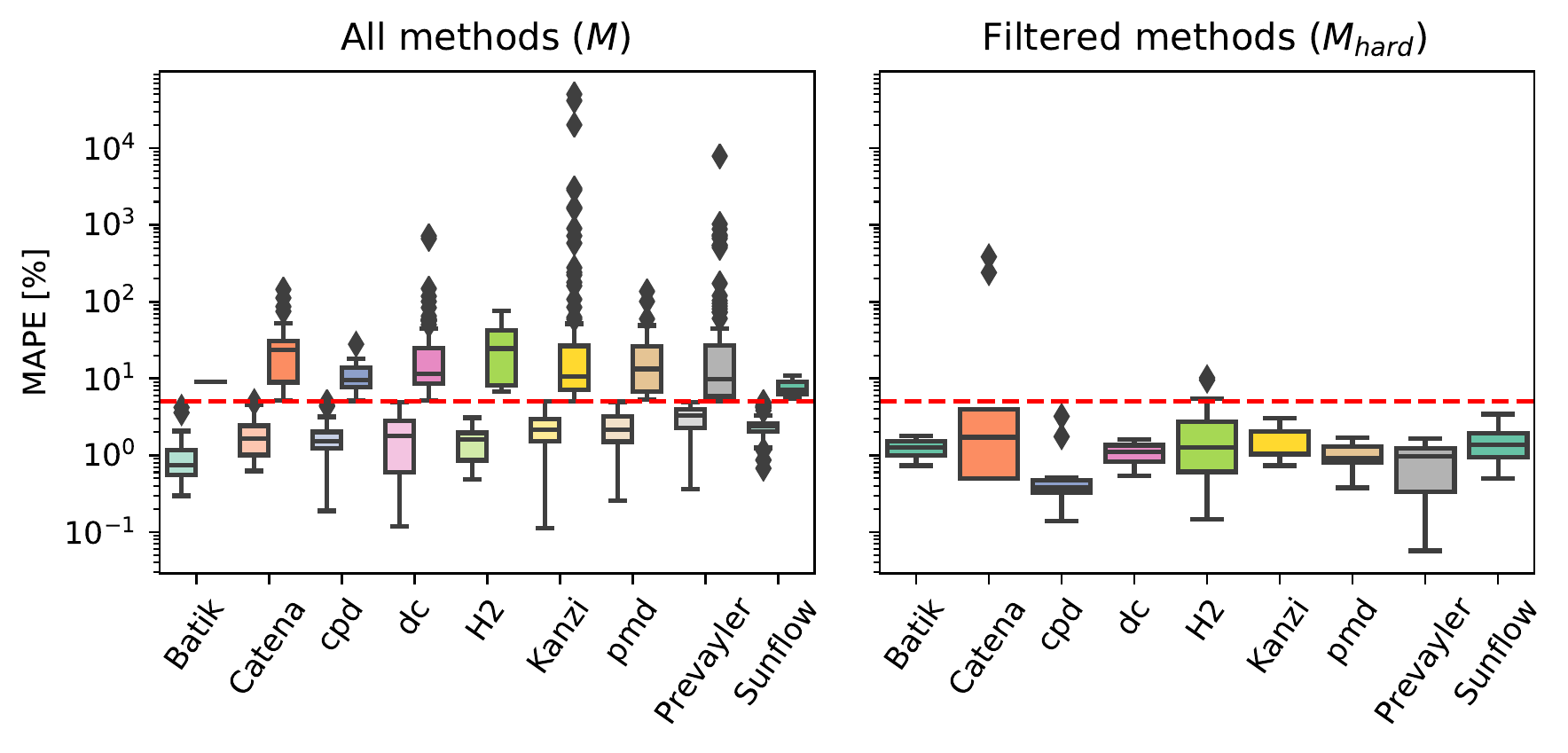}
    \caption{Error (MAPE) of method-level performance-influence models of all methods ($M$) and of the filtered methods ($M_{\mathit{hard}}$). Dashed red line denotes 5\,\% MAPE.}
    \label{fig:whitebox_accuracies}
\end{figure}

\begin{table}[t]
\caption{Method and correlation analysis. $|M|$ denotes the total number of methods and $|M_\mathit{hard}|$ the number of filtered methods. Configuration-wise linear correlation (LC) and rank correlation (RC) of black-box measurements (BB) using a coarse-grained profiler (CG) and a fine-grained profiler (FG).}
\centering

\begin{tabular}{l!{\qquad}rr!{\qquad}rrrr}
\toprule
    System &        $|M|$  & $|M_\mathit{hard}|$ & \multicolumn{2}{c}{BB vs.\ CG} & \multicolumn{2}{c}{BB vs.\ FG}
    \\\cmidrule(l){4-5}\cmidrule(l){6-7}
    & & & LC & RC & LC & RC \\
\midrule
    \textsc{Batik} &        122   &  9  & 0.82 &  0.89 &     0.76 &  0.87 \\
   \textsc{Catena} &        128   &  7  & 0.94 &  0.98 &     0.81 &  0.95 \\
      \textsc{cpd} &        125   &  11 & 0.25 &  0.55 &     0.57 &  0.76 \\
       \textsc{dc} &        435   &  3  & 0.51 &  0.91 &     0.53 &  0.96 \\
       \textsc{H2} &        238   &  43 & 0.32 &  0.42 &     0.12 &  0.34 \\
    \textsc{Kanzi} &        386   &  27 & 0.65 &  0.87 &     0.38 &  0.77 \\
      \textsc{pmd} &        218   &  5  & 0.32 &  0.72 &     0.18 &  0.38 \\
\textsc{Prevayler} &        166   &  7  & 0.94 &  0.97 &     0.84 &  0.88 \\
  \textsc{Sunflow} &        128   &  6  & 0.62 &  0.62 &     0.26 &  0.38 \\
\bottomrule
\end{tabular}
    \label{tab:hot-spots}

\end{table}

Figure~\ref{fig:whitebox_accuracies} summarizes the results regarding RQ$_1$. The plot on the left shows the prediction-error distribution of all method-level models based on data measured with the coarse-grained profiler. All methods that are below the dashed red marker (5\,\% prediction error) can be accurately learned based on a single profiling run per configuration, which makes up 84.8\,\% of all methods. Hence, the $\Pi_{\mathit{err}}$ condition of our filter selects about 15\,\% of the methods for step 2.
When using the additional conditions with respect to performance relevance ($\Pi_{\mathit{abs}}$ and $\Pi_{\mathit{rel}}$), we obtain a set comprising of about 6\,\% of all methods.
Table~\ref{tab:hot-spots} depicts for each system the set of filtered methods, denoted as $M_{\mathit{hard}}$.

Applying the second step, we are able to model nearly all of the remaining methods accurately as shown in Figure~\ref{fig:whitebox_accuracies} on the right-hand side. From a total number of $|M_{\mathit{hard}}|=118$ methods across all systems, only 5 methods cannot be learned accurately in step 2. A closer manual analysis of these methods revealed that they depend either heavily on thread-dependent file IO operations in the case of \textsc{H2} or nested loops (number of loops depends exponentially on an option) that copy an internal state array (width of array depends exponentially on a configuration option) in the case of \textsc{Catena}. 

\begin{mdframed}[style=mpdframe]
Answering RQ$_1$, the coarse-grained profiling step is able to learn models with a MAPE below 5\,\% for 84.8\,\% of all methods. Applying the fine-grained profiler in a second step, the MAPE is below 5\,\% for 95.8\,\% of the performance-relevant methods.
\end{mdframed}

\subsection{Tracing Option Influences ($RQ_2$)}





\paragraph{Operationalization}
To answer RQ$_2$, we focus on the importance of configuration options and interactions in white-box models. Specifically, we learn one random forest~\cite{geurts2006extremely}, consisting of 100 classification and regression trees, at system-wide level as well as one random forest for each method and extract all options and interactions.
For this analysis and without loss of generality, we concentrate on the two options (or interactions) per system that have the largest performance influence on the system determined by the black-box model (similar to the scenario sketched in Figure~\ref{fig:wbpim}). We aim at identifying the root cause for high influences by analyzing all performance-influence models at method level. Since there might be hundreds of methods per system, we analyze only performance-relevant methods, whose total sum can explain 80\,\% of the system performance. Having determined influential options and methods, we count those methods as root cause for which the configuration options (or interactions) we are interested in, have an influence that exceeds the measurement error. Furthermore, we sort the methods by the options' influences revealing, which method contributes most to an option's influence.

\paragraph{Results}

\begin{figure}
    \centering
    \includegraphics[width=0.98\linewidth]{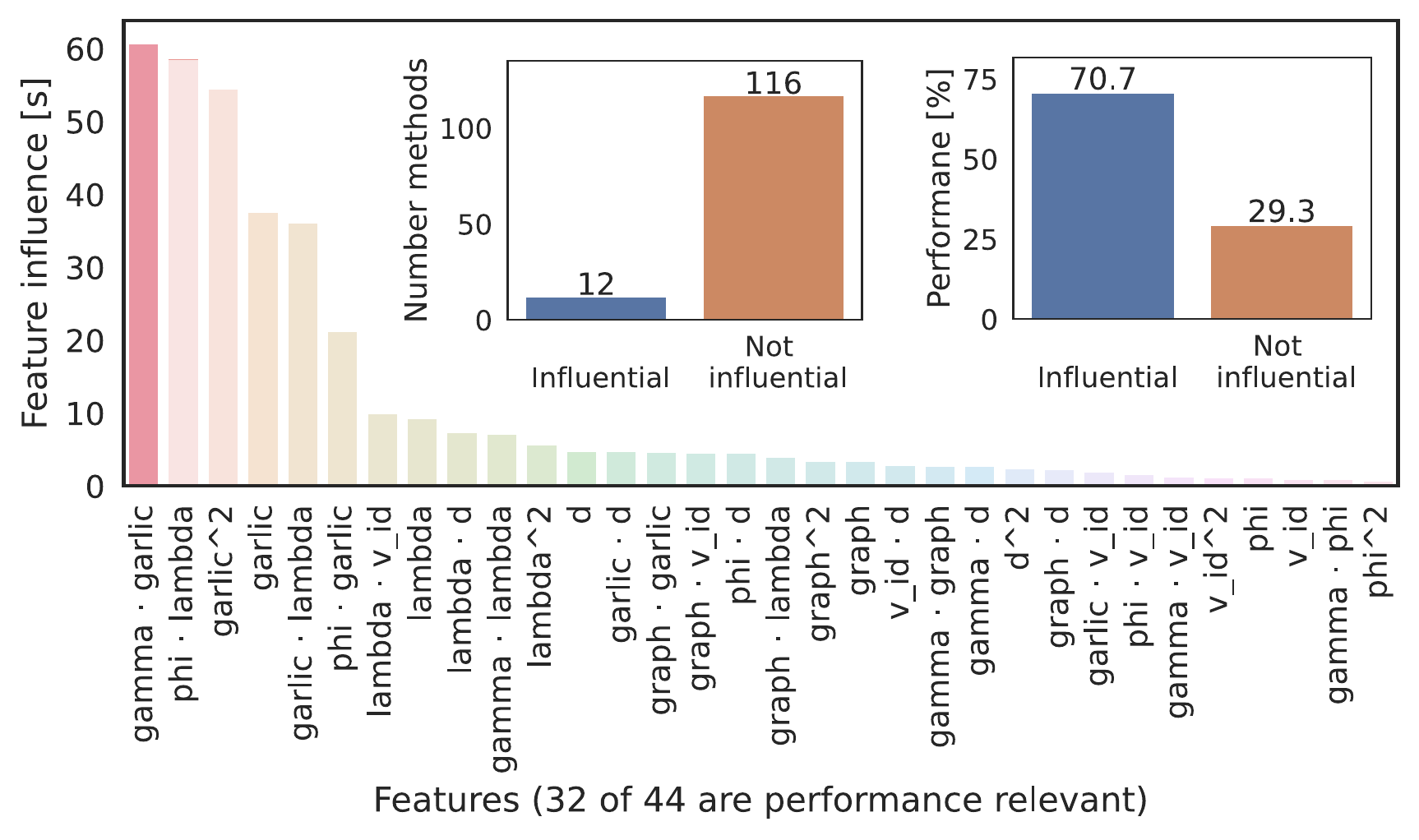}
    \caption{Overview of the influence of options and interactions on \textsc{Catena}'s performance. The background plot shows the distribution of all influential options and interactions (influence greater then the measurement error). The small inner plots focus on the interaction \textit{gamma $\cdot$ garlic}, which is the most influential option/interaction of the model. The left plot shows the number of methods that contribute and do not contribute to the interaction's performance. The right plot shows the performance portion of these methods.}
    \label{fig:eval:f_tracing}
\end{figure}


Tracing the influence of an option from the system level to the method level uncovers the cause of its influence. We present the results for \textsc{Catena} in Figure~\ref{fig:eval:f_tracing}. There are some configuration options and interactions that have an high influence on performance compared to the others; 12 options have no relevant influence. Focusing on an option of interest---the most important option in our example---reveals that only a small portion (12 out of 128) of the methods contribute to the option's system-wide performance influence. These 12 methods are responsible for more than 2/3 of the system's performance. This kind of information is not available in black-box models. It does not only help selecting important configuration options for guiding sampling and performance tuning, but also identifying the small set of methods that causes possible performance bugs, this way, facilitating performance bug detection of configurable systems.

\begin{mdframed}[style=mpdframe]
Answering RQ$_2$, white-box performance-influence models can successfully guide us to performance-relevant methods that are dependent on influential configuration options and interactions.
\end{mdframed}

\subsection{Profiled vs.\ Unprofiled Methods (RQ$_3$)}

\paragraph{Operationalization} To answer RQ$_3$, ideally we would need to compare the execution time between a profiled method and an unprofiled method.
Since we cannot know an unprofiled method's execution time, we use a proxy to infer the actual unprofiled method execution times. For this purpose, we consider the black-box execution time of a system as the aggregation of all true method's execution times. We compare this time against the aggregated (i.e., system-wide) predictions of white-box performance models for all methods. By repeating this process for all measured configurations, we approximate the relation of our estimates to the actual method execution times.

We use two different indicators to quantify the relation: Pearson's correlation coefficient and Spearman's rank correlation coefficient.
The former tests whether there is a linear dependency between the aggregated, system-wide execution time with and without profiling.
A high linear correlation would indicate that it is possible to infer the unprofiled execution time from the profiled execution time with a constant factor. This way, white-box performance influence models could even provide a precise prediction of method performance running in operation (without profiler).
A high linear correlation is unlikely, though, because the overhead during profiling increases while the number of profiled methods grows.
Spearman's rank correlation coefficient tests whether the order between profiled and unprofiled execution times is preserved.
A high rank correlation indicates that fast configurations measured without profiling stay fast, even if profiling is enabled. This would mean that our approach has accurately learned the relative influences of configuration options per method.

\paragraph{Results}
\begin{figure}
    \centering
    \includegraphics[width=0.98\linewidth]{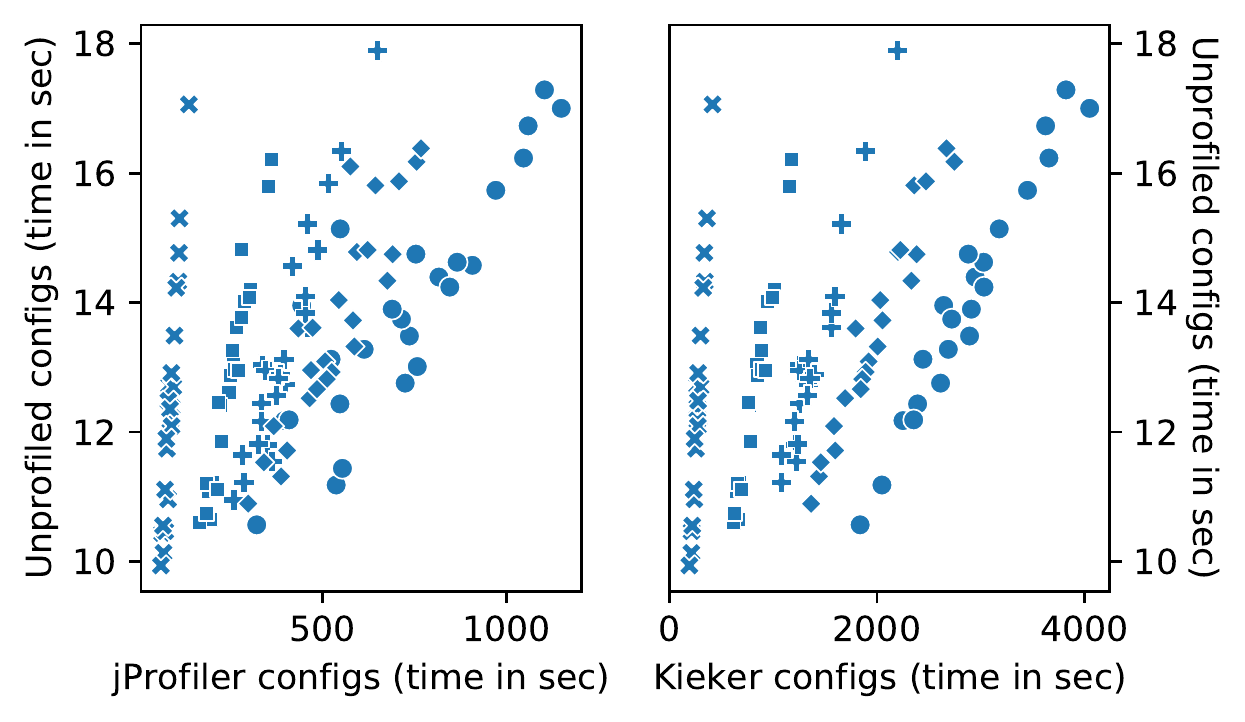}
    \caption{Configuration-wise execution time of \textsc{Sunflow}: black-box measurements vs.\ profiling. Left: using \textsc{jProfiler}, right: using \textsc{Kieker}. Different symbols visualized different numeric values of configuration option \textsc{samples}.}
    \label{fig:p_overhead:sunflow}
\end{figure}

In Table~\ref{tab:hot-spots}, we show the correlation between system-wide unprofiled execution time compared to using the coarse-grained profiler (BB vs.\ CG) and unprofiled execution time compared to using the fine-grained profiler (BB vs.\ FG). As expected, rank correlation is higher than linear correlation across all subject systems and for both profilers. That is, the fastest configurations remain the fastest independently of whether we use our learned models for predicting execution times or measuring execution time. This is good news as this property has been shown the main tuning objective for configurable systems~\cite{nair2017using}. Furthermore, we can see that different subject systems exhibit different correlations. Some systems, such as \textsc{Batik}, \textsc{Prevayler}, and \textsc{Catena}, exhibit even nearly perfect linear correlation. For them, the execution time depends strongly on the configuration for both types of experiments: measuring with a profiler and measuring the overall execution time of a program. There are also subject systems for which rank correlation is much higher than linear correlation: \textsc{Density Converter}, \textsc{Cpd}, and \textsc{Pmd}. There are two systems stand out with a generally low correlation: \textsc{Sunflow} and \textsc{H2}, which we analyze  next.

Figure~\ref{fig:p_overhead:sunflow} shows the dependency between unprofiled execution time and aggregated execution time measured with a profiler per configuration for \textsc{Sunflow}. We concentrate on the measurement overhead as the main cause for a low correlation. Interestingly, when highlighting configuration options, we observe a strong pattern for the overhead.
There are multiple groups that follow a linear trend, but with different slopes. The determining factor for this slope is the configured value of the numeric configuration option \textsc{samples}. This option is used as a seed for method \textsf{calculatePhotons} that compute the global illumination of the scene as part of the rendering process. By increasing the numeric value for this option, the overhead increases disproportionally. For \textsc{H2}, there is also a strong pattern, again, caused by a numeric configuration option (\textsc{analyzeAuto})~\footnote{More details on the supplementary Website.}. This is important for other studies in this area: Profiling overhead of configurable system depends on configuration options.


\begin{mdframed}[style=mpdframe]
Answering RQ$_3$, we observed a generally high rank correlation between system-wide profiled and unprofiled performance, demonstrating that the use of a profiler does not change the relative importance of configuration options and that white-box models are able to reveal performance-relevant methods. Some show even a linear correlation. A notable exception to this rule is a numeric option in \textsc{Sunflow}, which directly impacts the overhead introduced by the profiler.
\end{mdframed}

\subsection{Comparison to \textsc{ConfigCrusher}}
\label{sec:eval:cc}
Closest to our approach is \textsc{ConfigCrusher}~\cite{velez2019configcrusher}. The main difference to our approach is that \textsc{ConfigCrusher} relies on static taint analysis with the goal of determining which code regions are affected by which configuration options to weave measurement code at according statements.
Due to this conceptual difference, we face both qualitative as well as technical challenges that render a comparison infeasible. As explained in Section~\ref{rel_work:wb_ana}, there are three ways of propagating taints through a program: (A) taint every access to the data structure(s) that hold(s) the configuration options; (B) stop tainting at this point; and (C) rewrite the program, such that all options are stored in individual variables and are accordingly accessed across the code-base.

Variant C is infeasible in practice and also not in our case, as a substantial rewrite is not only impractical for larger systems such as \textsc{H2}, but this would also change the program structure such that a performance comparison has no longer a common ground.
For the purpose of comparison, we tried variant A first, but quickly run into timeouts and memory limitations, due to the inherent limitations of static code analysis. We communicated with \textsc{ConfigCrusher}'s main author, who confirmed our findings.

In a second attempt, we followed variant B. We again consulted the main author of \textsc{ConfigCrusher} for guidance to avoid introducing bias and setting up the subject system consistently with the original approach. As a result, we obtained reasonable taints for \textsc{Density-Converter}. For \textsc{H2}, the largest of our subject systems, we ran into memory overflows for all analyzed configuration options. The analysis of \textsc{Sunflow}, \textsc{Batik}, and \textsc{Prevayler} produced tainted code regions of size 1, which are basically useless. The reason was that the configuration options are immediately stored in a data structure, leading to a termination of the taint analysis. We provide all analysis log files at our supplementary Website. 


In addition to the limitations of the taint analysis, \textsc{ConfigCrusher} can handle only binary configuration options, whereas our approach can handle numeric options (as we do in our evaluation). Furthermore, \textsc{ConfigCrusher} is capable of tainting only single-threaded applications due to the underlying taint analysis. Our approach can produce performance models also for multi-threaded applications, such as \textsc{H2}.





\subsection{Threats to Validity}

The selection of the profiler represents a threat to construct validity. We mitigated this threat by a pre-study (not shown in the paper) where we evaluated several profilers. \textsc{JProfiler} is an industrial-strength profiler with low overhead, which turned out to be the best choice for the first phase. However, to obtain fine-grained performance data, we required more flexibility and opted for \textsc{Kieker}. A threat to internal validity arises from the measurement overhead introduced by the profiler. We reduce this threat by selecting a low-overhead profiler for measurement in the coarse-grained step and devoted a whole research question to analyze its influence.

The selection of subject systems threatens external validity. Although we cannot claim that we can learn white-box models accurately for all \textsc{Java} systems with proper profiling capabilities. Our results show that this is in principle possible for a large, industry-relevant branch of configurable software systems.

\section{conclusion}

We have proposed an approach to learn white-box performance-influence models at the method level, enabling tracing configuration effects from the system level to individual methods. Based on a pre-study on 3 software systems, we analyzed possible causes of performance variance of method execution times to design an integrated profiling and learning approach. We found that the majority of methods can be easily learned, as they either do not contribute much to the system's overall performance or do not contribute to configuration variance. Based on these insights, we have devised a two step approach in which we learn performance-influence models for all methods of a software system using a cheap, coarse-grained profiler in a first step, and filter inaccurate and relevant methods to be measured and learned again with an expensive, fine-grained profiler in a second step.
We found that, despite the overhead introduced by profiling, the correlation between profiled and unprofiled method execution time is high and that white-box models can accurately predict the profiled execution time. More importantly, we were able to show that performance models at the method level can be used to pinpoint the contribution of individual configuration options and interactions to individual methods, helping developers to chase configuration-related performance bugs or to focus performance testing on specific configurations.


\ifCLASSOPTIONcompsoc
  \section*{Acknowledgments}
\else
  \section*{Acknowledgment}
\fi

\revision{}{
Apel's work has been supported by the German Research Foundation (DFG) under the contract AP 206/11-1. Siegmund's work has been supported by the DFG under the contracts SI 2171/2 and SI 2171/3-1 and by the German Ministry of Education and Research (BMBF, 01IS19059A and 01IS18026B) by funding the competence center for Big Data and AI “ScaDS.AI Dresden/Leipzig”.
We thank our reviewers for their thoughtful comments. Especially, we thank Miguel Velez for his helpful comments on the specifics of the taint analysis and for supporting the set-up of ConfigCrusher for comparison.
}

\newpage



\bibliographystyle{IEEEtran}
\bibliography{monitoring}
%



\end{document}